\documentclass[twocolumn,showpacs,amsmath,amssymb,10pt,aps,superscriptaddress]{revtex4}
\usepackage{graphicx,color}
\usepackage{amsmath}
\usepackage{amssymb}
\usepackage{dsfont}
\usepackage{bbm}
\usepackage{color}

\usepackage[utf8]{inputenc}

\usepackage{amsfonts}
\usepackage{bm}

\newcommand{{\Cd}}{{\mathbb{C}^d}}

\newcommand{\BH}{\mathcal{B}(\mathcal{H})}

\def\oper{{\mathchoice{\rm 1\mskip-4mu l}{\rm 1\mskip-4mu l}
{\rm 1\mskip-4.5mu l}{\rm 1\mskip-5mu l}}}
\def\<{\langle}
\def\>{\rangle}
\newtheorem{thm}{Theorem}
\newtheorem{Lemma}{Lemma}

\newtheorem{Example}{Example}
\newtheorem{Proposition}{Proposition}

\newcommand{\beq}{\begin{equation}}
\newcommand{\eeq}{\end{equation}}
\newcommand{\bear}{\begin{eqnarray}}
\newcommand{\ear}{\end{eqnarray}}
\newcommand{\bdm}{\begin{displaymath}}
\newcommand{\edm}{\end{displaymath}}

\def\wb{{\mathbf{w}}}
\def\s{\sigma}

\def\w{\omega}

\def\a{\alpha}
\def\b{\beta}
\def\g{\gamma}

\def\l{\lambda}

\begin{document}

\title{%On Kadison-Schwarz divisibility and  quantum non-Markovianity\\
Dissipative generators, divisible dynamical maps and Kadison-Schwarz inequality}
\author{Dariusz Chru\'sci\'nski}
\affiliation{Institute of Physics, Faculty of Physics, Astronomy and Informatics, Nicolaus Copernicus University, Grudzi\c{a}dzka 5/7, 87--100 Toru\'n, Poland}

\author{Farrukh Mukhamedov}
\affiliation{Department of Mathematical Sciences, Collage of Science, United Arab Emirates University, Al Ain, 155511, Abu Dhabii, UAE}

\begin{abstract}
We introduce a concept of Kadison-Schwarz divisible dynamical maps. It turns out that it is a natural generalization of the well known CP-divisibility which characterizes quantum Markovian evolution. It is proved that Kadison-Schwarz divisible maps are fully characterized in terms of time-local dissipative generators. Simple qubit evolution illustrates the concept.
\end{abstract}


\maketitle

\section{Introduction}

Evolution of a quantum system is represented by a family of quantum channels $\Lambda_t : \BH \to \BH$  ($t \geq 0$) such that $\Lambda_0 = {\rm id}$ (identity map). In what follows $\BH$ denotes an algebra of bounded linear operators acting in the Hilbert space $\mathcal{H}$ (actually, in this paper we consider only finite dimensional $\mathcal{H}$). Such family is usually called a dynamical map. For isolated system the dynamical map has a well known structure $\Lambda_t(\rho) = U_t \rho U_t^\dagger$, where $U_t = e^{-iHt}$ and $H$ denotes the Hamiltonian of the (closed) system (we keep $\hbar=1$). For an open quantum system one often considers a dynamical semigroup governed by  the Markovian master equation \cite{Breuer,RivasHuelga}

\begin{equation}\label{ME}
  \dot{\Lambda}_t = \mathcal{L} \Lambda_t ,
\end{equation}
where the generator $\mathcal{L} : \BH \to \BH$ is given by the celebrated Gorini-Kossakowski-Sudarshan-Lindblad formula \cite{GKLS,Lindblad}

\begin{equation}\label{GKSL}
  \mathcal{L}(\rho) = -i[H,\rho] + \sum_k \gamma_k \left( V_k \rho V_k^\dagger - \frac 12 \{ V_k^\dagger V_k,\rho\} \right) ,
\end{equation}
with positive rates $\gamma_k > 0$. This structure guarantees that the solution $\Lambda_t = e^{t \mathcal{L}}$ defines a legitimate dynamical map -- completely positive and trace preserving (CPTP). To go beyond dynamical semigroup one considers master equation (\ref{ME}) with time dependent generator $\mathcal{L}_t$. It has exactly the same form as in (\ref{GKSL}) with time dependent $H(t)$, $V_k(t)$ and $\gamma_k(t)$. The formal solution for $\Lambda_t$ reads

\begin{equation}\label{T}
  \Lambda_t = \mathcal{T}_{\leftarrow} \exp\left( \int_0^t \mathcal{L}_\tau d\tau \right) ,
\end{equation}
where $\mathcal{T}_{\leftarrow}$ stands for chronological operator. In this case, however, we do not known necessary and sufficient conditions for $\mathcal{L}_t$ which guarantee that (\ref{T}) is CPTP for all $t > 0$. Time dependent generators are recently analyzed in connection to quantum non-Markovian evolution \cite{NM1,NM2,NM3,NM4}. Recall, that a dynamical map $\Lambda_t$ is called divisible if

\begin{equation}\label{}
  \Lambda_t = V_{t,s} \Lambda_s , \ \ t > s ,
\end{equation}
with a family of `propagators' $V_{t,s} : \BH \to \BH$. For invertible maps such propagator always exists and it is given by $V_{t,s}= \Lambda_t \Lambda_s^{-1}$. Maps which are not invertible require special treatment \cite{PRL-2018,Sagnik} (for a recent review of various concepts of divisibility for quantum channels and dynamical maps see \cite{Mario}).  Now, being divisible one calls $\Lambda_t$ P-divisible if $V_{t,s}$ is positive and trace-preserving, and CP-divisible if $V_{t,s}$ is CPTP. Following \cite{RHP} one calls the evolution represented by $\Lambda_t$ Markovian if $\Lambda_t$ is CP-divisible. Actually, for invertible maps CP-divisibility is fully controlled  by the properties of time-local generator $\mathcal{L}_t$, that is, all time-dependent rates satisfy $\gamma_k(t) \geq 0$. Authors of \cite{BLP} proposed another approach based on distinguishability of quantum states: $\Lambda_t$ is Markovian if for any pair of initial states $\rho$ and $\sigma$ one has

\begin{equation}\label{BLP}
  \frac{d}{dt} \| \Lambda_t(\rho) - \Lambda_t(\sigma) \|_1 \leq 0 ,
\end{equation}
where $\| X \|_1 = {\rm Tr}\sqrt{X^\dagger X}$. Interestingly, for invertible maps P-divisibility is equivalent to

\begin{equation}\label{}
  \frac{d}{dt} \| \Lambda_t(X) \|_1 \leq 0 ,
\end{equation}
for all $X^\dagger =X$. Hence, CP-divisibility implies P-divisibility and this implies BLP condition (\ref{BLP}).

In this paper we introduce another notion of divisibility based on the Kadison-Schwarz (KS) inequality. Let us recall that a linear map $\Phi : \BH \to \BH$ is trace-preserving iff its dual map $\Phi^\sharp$ is unital. $\Phi^\sharp$ is defined by ${\rm Tr}(X \Phi(Y)) = {\rm Tr}(\Phi^\sharp(X) Y)$ for all $X,Y \in \BH$. Hence, if $   \Phi(X) = \sum_k \lambda_k A_k X B_k^\dagger$, then $  \Phi^\sharp(X) = \sum_k \lambda^*_k B_k X A_k^\dagger$. In particular, if $\Lambda_t(\rho) = U_t \rho U^\dagger_t$ represents Schr\"odinger evolution of the density operator, then $\Lambda^\sharp_t(X) = U^\dagger_t X U_t$ represents Heisenberg  evolution of the observable $X$. Introducing a completely positive map $\Phi(\rho) = \sum_k \gamma_k  V_k \rho V_k^\dagger$, the GKSL generator (\ref{GKSL}) can be rewritten in a compact form as follows

\begin{equation}\label{LI}
  \mathcal{L}(\rho) = -i[H,\rho] + \Phi(X) - \frac 12 \{ \Phi^\sharp(\oper),X \}  .
\end{equation}
Now, a unital map $\Phi : \BH \to \BH$ satisfies Kadison-Schwarz (KS) inequality \cite{Paulsen,Stormer,Kadison,Bhatia} if
\begin{equation}\label{KS0}
  \Phi(XX^\dagger) \geq \Phi(X) \Phi(X^\dagger) ,
\end{equation}
for all $X \in \BH$. We say that dynamical map $\Lambda_t$ is KS-divisible if the propagator $V^\sharp_{t,s}$ satisfies (\ref{KS0}). In this paper we analyze this concept and relate it to P- and CP-divisibility. Interestingly, for invertible maps KS-divisibility is fully controlled by the property of $\mathcal{L}_t$ called dissipativity \cite{Lindblad}. Finally, we illustrate KS-divisibility by simple example of qubit evolution.

\section{Kadison-Schwarz maps}

Celebrated Cauchy-Schwarz inequality found a lot of applications in mathematics, physics and engineering. It states that for any $x,y \in \mathcal{H}$ one has $|\langle x|y\rangle|^2 \leq \langle x|x\rangle \langle y|y\rangle$. Kadison found elegant generalization of this inequality for linear maps in operator algebras \cite{Kadison,Paulsen,Stormer,Bhatia}: a linear map $\Phi : \BH \to \BH$ is positive if for any $A\geq 0$ one has $\Phi(A) \geq 0$. Equivalently, for any $X \in \BH$ one has $\Phi(X^\dagger X) \geq 0$. A linear map is unital if $\Phi(\oper)=\oper$, with $\oper$ being an identity operator in $\mathcal{H}$. Now,  a unital map $\Phi : \BH \to \BH$ satisfies Kadison-Schwarz (KS) inequality \cite{Paulsen,Kadison,Bhatia} if
\begin{equation}\label{KS}
  \Phi(XX^\dagger) \geq \Phi(X) \Phi(X^\dagger) ,
\end{equation}
for all $X \in \BH$. It immediately follows from (\ref{KS}) that Kadison-Schwarz map is positive. However, the converse needs not be true. An example of a positive unital map which is not KS is provided by the transposition map. Indeed, for $d=2$ taking $X=|1\rangle \langle 2|$ one finds

$$   T(X^\dagger X ) = |1\rangle \langle 1| , \ \ T(X^\dagger)T(X) = |2\rangle \langle 2| , $$
and hence (\ref{KS}) is violated. Interestingly, any unital positive map satisfies (\ref{KS}) but for normal operators (i.e. $X^\dagger X = X X^\dagger$). Denote by $M_k(\mathbb{C}$ linear space of $k\times k$ complex matrices. Recall, that a linear map $\Phi$ is $k$-positive if the extended map

$$   {\rm id}_k : M_k(\mathbb{C}) \otimes \BH \to M_k(\mathbb{C}) \otimes \BH $$
is positive (${\rm id}_k$ denotes an identity map). A map which is $k$-positive for $k=1,2,\ldots$ is called completely positive (CP). If the dimension of $\mathcal{H}$ is `$d$', then complete positivity is equivalent to $d$-positivity. Among unital maps one has the following hierarchy

\begin{equation}\label{PPPP}
   {\rm CP} ={\rm P}_d \subset {\rm P}_{d-1} \subset \ldots \subset {\rm P}_2 \subset {\rm KS} \subset {\rm P}_1  ,
\end{equation}
where ${\rm P}_k$ denotes $k$-positive unital maps.

If $\Phi$ and $\Psi$ are KS, then $\Phi \Psi$ is KS as well. Moreover, the convex combination $\lambda \Phi + (1-\lambda) \Psi$ is again KS \cite{Farrukh}.

Actually, the concept of unital KS maps may be generalized for maps which are not unital. Consider a map $\Phi : \BH \to \BH$ such that $\Phi(\oper) > 0$, and define

\begin{equation}\label{}
  \Psi(X) = \Phi(\oper)^{-\frac 12} \Phi(X) \Phi(\oper)^{-\frac 12} .
\end{equation}
Clearly, one has $\Psi(\oper)=\oper$. Now, if $\Psi$ satisfies KS condition (\ref{KS}), then

\begin{equation}\label{KS-2}
  \Phi(XX^\dagger) \geq \Phi(X)  \Phi(\oper)^{-1} \Phi(X^\dagger) .
\end{equation}

\begin{Example} Consider a qubit map

\begin{equation}\label{}
  \Phi = p_1 \Phi_1 + p_2 \Phi_2 + p_3 \Phi_3 ,
\end{equation}
where $p_1+p_2+p_3=1$, and

$$   \Phi_k(X) = \frac 12( \sigma_k X \sigma_k + X) , $$
with $\sigma_k$ being Pauli matrices. Note, that

$$   \Phi_k(X) = \sum_{\mu=1}^{2} P^{(\mu)}_k X P^{(\mu)}_k , $$
and $P^{(\mu)}_k$ are eigen-projectors of $\sigma_k$. It is clear that $\Phi$ is unital. It is CP iff $p_k \geq 0$. Note, that

$$   \Phi(\sigma_k) = p_k \sigma_k , $$
and hence one easily finds that $\Phi$ is positive iff $|p_k|\leq 1$. For example the map

$$   \Phi = \Phi_1 +  \Phi_2 - \Phi_3 , $$
is positive (but of course not CP). This map is not KS. Indeed, taking $X = |1\rangle \langle 2|$ one gets

$$    \Phi(X^\dagger X) = |1\rangle \langle 1| , \ \  \Phi(X^\dagger)\Phi(X) =  |2\rangle \langle 2| . $$
It is shown in the Appendix that if

\begin{equation}\label{ppp}
 p_1^2 + p_2^2 + p_3^2 \leq 1 + 2 p_1p_2p_3 ,
\end{equation}
then $\Phi$ is KS (cf. Figure 1). Interestingly, the constraint (\ref{ppp}) provides good approximation for a set of CP maps (cf. Figure 1). For example all three vertices of the inscribed triangle satisfy (\ref{ppp}) with equality.

\begin{figure*}
  \includegraphics[scale=0.4]{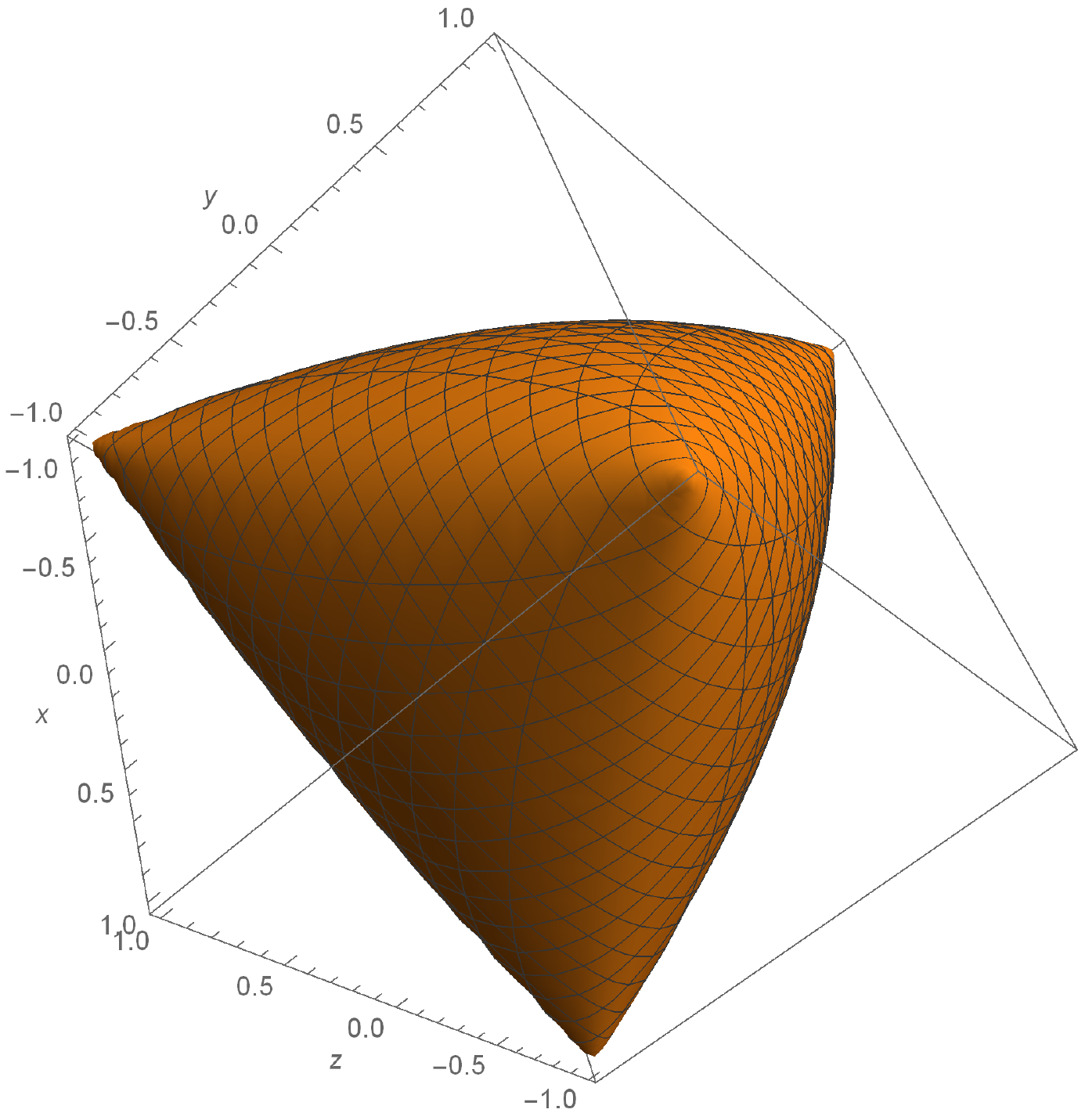} \includegraphics[scale=0.4]{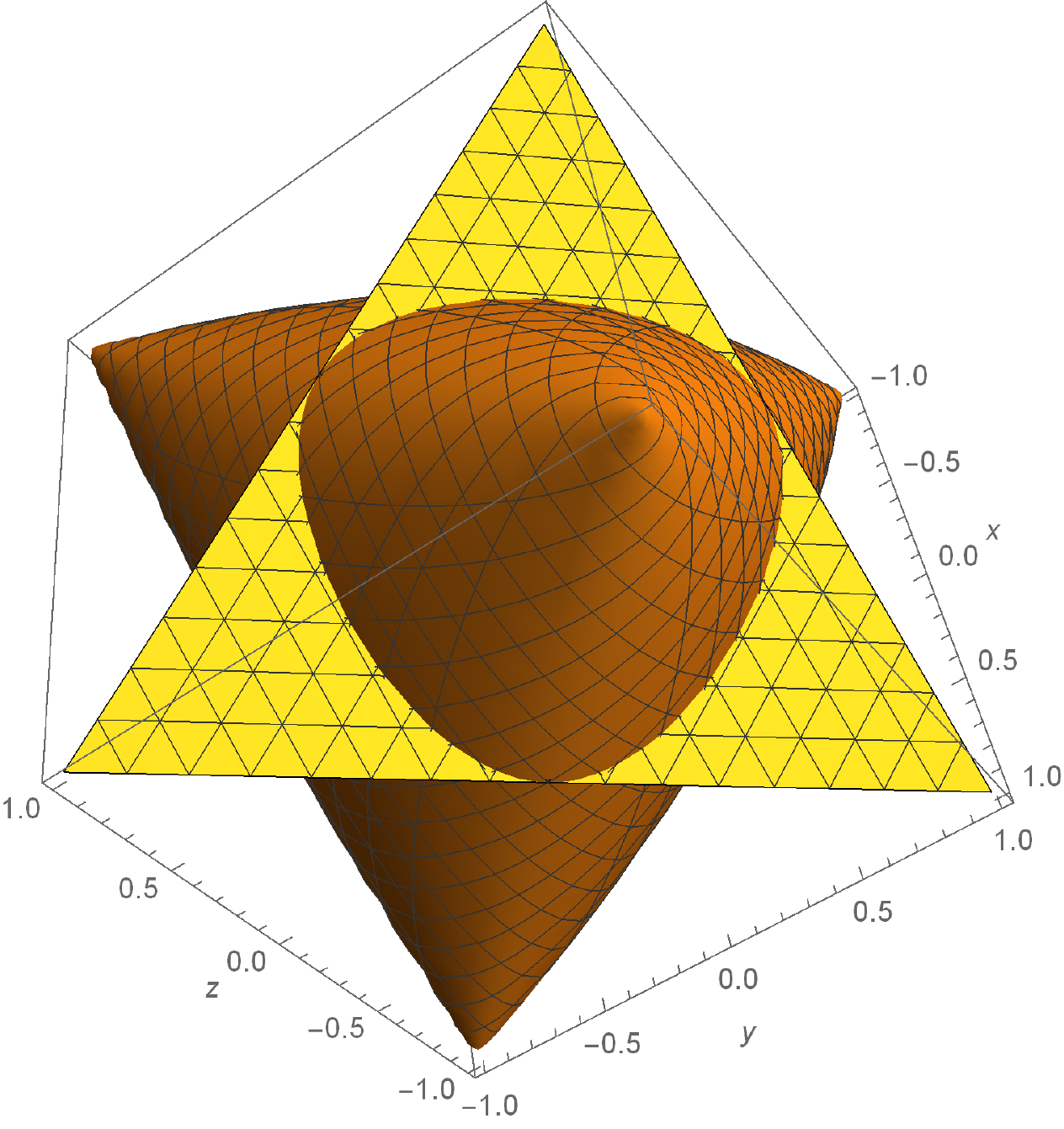}  \includegraphics[scale=0.4]{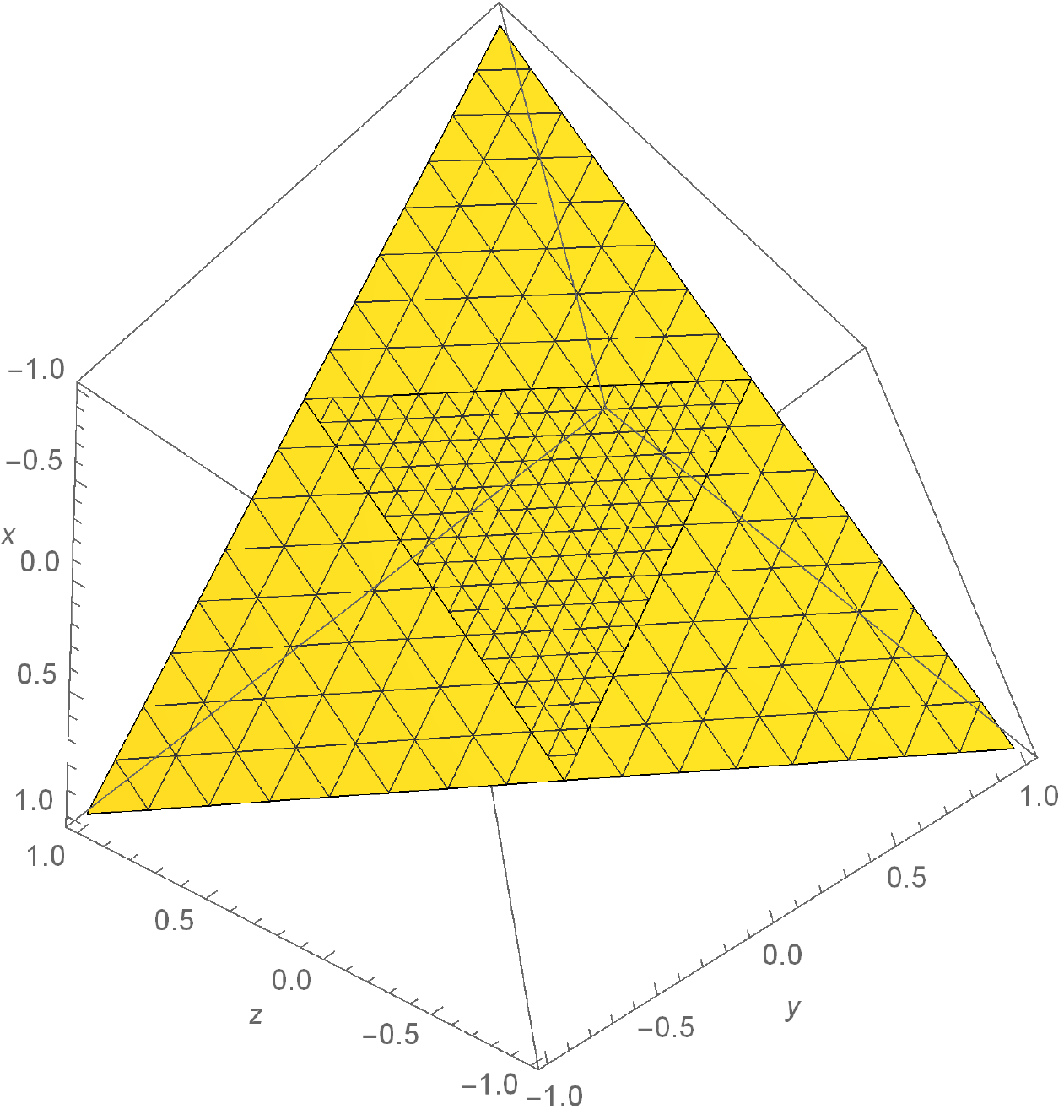}
  \caption{Left panel: the convex body satisfying (\ref{ppp}). Middle panel: the intersection with the plane $p_1+p_2+p_3=1$. The yellow triangle with vertices $(1,1,-1)$, $(1,-1,1)$ and $(-1,1,1)$ corresponds to positive maps. Right panel: inscribed triangle of CP maps with vertices $(1,0,0)$, $(0,1,0)$ and $(0,0,1)$. }
 \end{figure*}

\end{Example}

\section{KS-divisibility}

It was already observed by Lindblad \cite{Lindblad} that $\Lambda^\sharp_t = e^{t \mathcal{L}^\sharp}$ is KS iff $\mathcal{L}^\sharp$ is dissipative, that is,

\begin{equation}\label{}
  \mathcal{L}^\sharp(X^\dagger X) \geq \mathcal{L}^\sharp(X^\dagger) X + X^\dagger\mathcal{L}^\sharp(X) ,
\end{equation}
for all $X \in \BH$. Any dissipative generator has the following structure

\begin{equation}\label{LII}
  \mathcal{L}^\sharp(X) = i[H,X] + \Phi(X) - \frac 12 \{ \Phi(\oper),X \}  ,
\end{equation}
where the map $\Phi : \BH \to \BH$ satisfies the following condition

\begin{equation}\label{QKS}
   \Phi(X^\dagger X) \geq \Phi(X^\dagger) X + X^\dagger\Phi(X) - X^\dagger \Phi(\oper) X ,
\end{equation}
that is, it has exactly the same structure as (\ref{LI}) but the CP map $\Phi$ is replaced by the map satisfying (\ref{QKS}) (note that (\ref{LI}) is represented in the Schr\"odinger picture whereas (\ref{LII}) in the Heisenberg picture). Actually condition (\ref{QKS}) is weaker than generalized KS condition (\ref{KS-2}). One has

\begin{Proposition} Any $\Phi$ satisfying (\ref{KS-2}) satisfies (\ref{QKS}).
\end{Proposition}
For the proof see Appendix.

Consider now a dynamical map $\Lambda_t$ satisfying time-local master equation $\dot{\Lambda}_t = \mathcal{L}_t \Lambda_t$, that is, $\Lambda_t$ is represented as in (\ref{T}).

\begin{thm} If $\Lambda_t$ is invertible, then

\begin{itemize}
  \item it is KS-divisible if and only if $\mathcal{L}^\sharp_t$ is dissipative,

  \item it is CP-divisible if and only if $\mathcal{L}^\sharp_t$ is completely dissipative
\end{itemize}
for all $t \geq 0$.

\end{thm}
Proof: the existence of $V_{t,s}$ follows from invertibility of $\Lambda_t$, that is, $V_{t,s} = \Lambda_t \Lambda_s^{-1}$. One has

  \begin{equation}\label{V-ts}
  V_{t,s}^\sharp = \mathcal{T}_{\rightarrow} \exp\left( \int_s^t \mathcal{L}^\sharp_\tau d\tau \right) ,
\end{equation}
where now $\mathcal{T}_{\rightarrow}$ stands for anti-chronological operator. Now, if $\mathcal{L}^\sharp_t$ is dissipative, then $V_{t,s}^\sharp$ is unital KS. If $V_{t,s}^\sharp$ is KS for any $t > s$, then for $\epsilon\to 0+$ one has $V^\sharp_{t+\epsilon,t}\to e^{\epsilon \mathcal{L}_t^\sharp}$ which implies that  $\mathcal{L}^\sharp_t$ is dissipative. \hfill $\Box$

\section{Examples: KS-divisible qubit dynamical maps}

In this section we consider several simple examples of qubit evolution. In this case the hierarchy (\ref{PPPP}) reduces to

\begin{equation}\label{}
   {\rm CP} ={\rm P}_2 \subset {\rm KS} \subset {\rm P}_1  ,
\end{equation}
that is, KS maps interpolate between CP and positive maps.

\begin{Example}[Qubit dephasing] For a qubit dephasing governed by
\begin{equation}\label{}
  \mathcal{L}_t(\rho) = \gamma(t) (\sigma_3 \rho \sigma_3 - \rho) ,
\end{equation}
P-, KS-, and CP-divisibility coincide and they are equivalent to $\gamma(t) \geq 0$. Note, that in this case one has $\mathcal{L}^\sharp_t = \mathcal{L}_t$.
\end{Example}

\begin{Example}[Amplitude damping channel]
The evolution of amplitude-damped qubit is
governed by a single function $G(t)$
\begin{equation}\label{}
  \Lambda_t(\rho) = \left( \begin{array}{cc} \rho_{11} + (1-|G(t)|^2)\rho_{22} & G(t) \rho_{12} \\ G^*(t) \rho_{21} & |G(t)|^2 \rho_{22} \end{array} \right) ,
\end{equation}
where the function $G(t)$ depends on the form of the reservoir spectral density $J(\omega)$ \cite{Breuer}.
This evolution is generated by the following time-local generator
\begin{equation*}\label{}
  \mathcal{L}_t(\rho) = - i[H(t),\rho] + \gamma(t)( \sigma_- \rho \sigma_+ - \frac 12 \{ \sigma_+\sigma_-,\rho\} ) ,
\end{equation*}
where $\sigma_\pm$ are the spin lowering and rising operators, $H(t)= \frac{\omega(t)}{2}\sigma_+\sigma_-$,  together with
$\omega(t) = -2{\rm Im} \frac{\dot{G}(t)}{G(t)}$, and $\gamma(t) = -2{\rm Re} \frac{\dot{G}(t)}{G(t)}$. . Again in this case P-, KS-, and CP-divisibility coincide and they are equivalent to $\gamma(t) \geq 0$.
\end{Example}

\begin{Example}[Pauli channel]  Consider the qubit evolution governed by the following time-local generator
\begin{equation}\label{Pauli}
  \mathcal{L}_t(\rho) = \frac 12 \sum_{k=1}^3  \gamma_k(t)( \sigma_k \rho \sigma_k - \rho) ,
\end{equation}
which leads to the following dynamical map (time-dependent Pauli channel):

\begin{equation}\label{Pauli}
\Lambda_t(\rho) =  \sum_{\alpha=1}^3  p_\alpha(t) \sigma_\alpha \rho \sigma_\alpha ,
\end{equation}
where

$$   \left( \begin{array}{c} p_0(t) \\  p_1(t) \\  p_2(t) \\  p_3(t) \end{array} \right) =    \frac 14 \left( \begin{array}{rrrr} 1 & 1 & 1 & 1 \\  1 & 1 & -1 & -1  \\  1 & -1 & 1 & -1  \\  1 & -1 & -1 & 1 \end{array} \right) \left( \begin{array}{c} 1 \\  \lambda_1(t) \\ \lambda_2(t) \\  \lambda_3(t) \end{array} \right) , $$
and

$$ \lambda_1(t) = e^{- \Gamma_2(t) - \Gamma_3(t) } \ \ + {\rm cyc. permutations} , $$
with $\Gamma_k(t) = \int_0^t \gamma_k(\tau) d\tau$. It was shown \cite{Filip-PRA} that P-divisibility is equivalent to the following conditions:

\begin{equation}\label{gg}
  \gamma_i(t) +  \gamma_j(t) \geq 0 ,   \ \ \ i \neq j  .
\end{equation}
Now, it is shown in the Appendix that KS-divisibility is equivalent to the following the stronger conditions:

\begin{equation}\label{g2g}
  \gamma_i(t) +  2 \gamma_j(t) \geq 0 ,  \ \ \ i \neq j  .
\end{equation}
In \cite{Erika} authors considered so called eternally non-Markovian evolution corresponding to

\begin{equation}\label{erika}
  \gamma_1(t)=\gamma_2(t)=1 , \ \ \gamma_3(t) = - {\rm tanh}\, t .
\end{equation}
It gives
\begin{eqnarray*}
% \nonumber % Remove numbering (before each equation)
  p_0(t) &=& \frac 12 (1+ e^{-2t}) ,  \\
  p_1(t) &=& p_2(t) = \frac 14 (1- e^{-2t}) , \\
  p_3(t) &=& 0 .
\end{eqnarray*}
Clearly, the map $\Lambda_t$ is CPTP and P-divisible. Now, let us consider a simple modification

\begin{equation}\label{}
  \gamma_1(t)=\gamma_2(t)=1 , \ \ \gamma_3(t) = - \frac 12 {\rm tanh}\, t .
\end{equation}
It gives

\begin{eqnarray*}
% \nonumber % Remove numbering (before each equation)
  p_0(t) &=& \frac 14 (1+  2 e^{-t} \sqrt{\cosh t} + e^{-2t}) ,  \\
  p_1(t) &=& p_2(t) = \frac 14 (1- e^{-2t}) , \\
  p_3(t) &=& \frac 14 (1- 2 e^{-t} \sqrt{\cosh t} + e^{-2t}) .
\end{eqnarray*}
Again, the map $\Lambda_t$ is CPTP due to the fact that $p_\alpha(t) \geq 0$. Indeed, one finds

$$   p_3(t) = \frac{e^{-t}}{2}( \cosh t - \sqrt{\cosh t} ) \geq 0 , $$
due to $\cosh t \geq 1$.
% (cf. Fig. 2 for the plot of $p_3(t)$).
%\begin{figure*}
%  \includegraphics[scale=0.4]{p3-plot.pdf}
%  \caption{The evolution of $p_3(t)$. }
% \end{figure*}
Hence, it provides an example of KS-divisible qubit evolution since conditions (\ref{g2g}) are trivially satisfied. Clearly, the evolution governed by (\ref{erika}) is P-divisible but not KS-divisible.
\end{Example}

\section{Conclusions}

In this paper we introduced the concept of KS-divisibility which is based on the Kadison-Schwarz inequality (\ref{KS}). This concept interpolates between CP-divisibility(often assumed as a definition of Markovianity \cite{RHP}) and P-divisibility (which is closely related to the well known notion of information flow \cite{BLP}). Any CP-divisible map in KS-divisible, and any KS-divisible map is P-divisible and hence does not display information backflow. For dynamical maps satisfying time-local master equation with time-dependent generator $\mathcal{L}_t$ KS-divisibility is fully controlled by the property of the generator $\mathcal{L}_t^\sharp$ (Heisenberg picture), that is, the maps is KS-divisible if and only if $\mathcal{L}_t^\sharp$ is dissipative. This concept is illustrated by several examples of qubit evolution. Interestingly for the evolution governed by well known generator $\mathcal{L}_t(\rho) = \frac 12 \sum_k \gamma_k(t) (\sigma_k \rho\sigma_k - \rho)$ we found necessary and sufficient conditions for dissipativity: $\gamma_i(t) + 2\gamma_j(t) \geq 0$ for $i\neq j$. I shows that so called eternally non-Markovian evolution proposed in \cite{Erika} is P-divisible but not KS-divisible. However, we proposed a simple modification which is again eternally non-Markovian (one of the rate is always negative) but displays KS-divisibility. Actually, condition  $\gamma_i(t) + 2\gamma_j(t) \geq 0$ for $i\neq j$ has a clear physical interpretation: note that the initial Bloch vector $\mathbf{x}=(x_1,x_2,x_3)$ evolves according to

\begin{equation}\label{}
  \mathbf{x}(t) = ( \lambda_1(t) x_1, \lambda_2(t) x_2, \lambda_3(t) x_3) , 
\end{equation}
where $\lambda_k(t) = \exp(-\int_0^t 1/T_k(\tau) d\tau)$, and the local relaxation times read

$$    T_i(t) = \frac{1}{\gamma_j(t)+\gamma_k(t)} , $$
for mutually different $i,j,k$. Now, assuming that $\gamma_3(t) < 0$, one finds the following constraints

$$  T_1(t),T_2(t) \leq \frac{1}{|\gamma_3(t)|}, \ \ T_3(t) \leq \frac{1}{4|\gamma_3(t)|} . $$
Note, that if the map is only P-divisible one has  $T_3(t) \leq \frac{1}{2|\gamma_3(t)|}$ and no additional constraints for $T_1(t)$ and $T_2(t)$. This shows that these two concepts of divisibility have different physical flavour. 

It would be interesting to investigate KS-divisibility for higher dimensional systems. 

\section*{Acknowledgments}

DC was supported by the National Science Center project No 2018/30/A/ST2/00837.

\appendix

\section{Condition (\ref{ppp}) for KS map}

We note that every matrix $X\in\textit{M}_2(\mathbb{C})$ can be written in this basis as
$X = w_0 \oper + \wb\cdot\s$ with $w_0\in\mathbb{C}, \wb=(w_1,w_2,w_3)\in\mathbb{C}^3,$ here by $\wb\cdot\s$ we mean the following $$\wb\cdot\s=w_1\s_1+w_2\s_2+w_3\s_3.$$

\begin{widetext}

One finds
\begin{equation}\label{qks3}
\Phi(w_0 \oper +\wb\cdot\s)=w_0\oper + T\wb\cdot\s ,
\end{equation}
where the $3\times 3$ matrix $T_{ij}$ reads $T_{ij} = p_i \delta_{ij}$.
%$$
%T(\w_1,\w_2,\w_3)=(p_1\w_1, p_2\w_2,p_3\w_3), \ \  \ \ \textrm{where} \ \ \ \wb=(\w_1\,\w_2,\w_3)
%$$
Taking into account  a result of \cite{Farrukh} the KS conditions yields

\begin{eqnarray}\label{q34}
\bigg(A|w_2\overline{w}_3-w_3\overline{w}_2|^2&+&B|w_1\overline{w}_3-w_3\overline{w}_1|^2 +C|w_1\overline{w}_2-w_2\overline{w}_1|^2\bigg)^{1/2} \leq \a|w_1|^2+\b |w_2|^2+\g |w_3|^2
\end{eqnarray}
where
\begin{eqnarray}\label{abc}
&&\alpha=|1-p_1^2|,  \ \ \ \beta=|1-p_2^2|, \ \ \ \gamma=|1-p_3^2|\\
\label{abc1}
&&A=|p_1-p_2p_3|^2, \ \ \ B=|p_2-p_1p_3|^2, \ \ \ C=|p_3-p_1p_2|^2.
\end{eqnarray}
Let us assume $\|\wb\|=1$. Hence, taking into account $w_1=r_1e^{i\a_1}$, $w_2=r_2 e^{i\a_2}$, $w_3=r_3 e^{i\a_3}$, the inequality \eqref{q34} reduces to

\begin{eqnarray}\label{q35}
2\bigg(A r_2^2r_3^2\sin^2\theta_1 +B r_1^2r_3^2\sin^2\theta_2 +C r_1^2r_2^2\sin^2\theta_3\bigg)^{1/2}\leq \a r_1^2+\b r_2^2+\g r_3^2
\end{eqnarray}
where $\theta_1+\theta_2+\theta_3=2\pi$. Clearly, the last inequality is satisfied if one has
\begin{eqnarray}\label{q36}
2\bigg(A r_2^2r_3^2+B r_1^2r_3^2+C r_1^2r_2^2\bigg)^{1/2}\leq \a r_1^2+\b r_2^2+\g r_3^2
\end{eqnarray}
Introducing $x:=r_1^2, y:=r_2^2, z:=r_3^2$, the last one is equivalent to
\begin{eqnarray}\label{q37}
2\bigg( A yz+B xz+C xy \bigg)^{1/2}\leq \a x+\b y+\g z
\end{eqnarray}
for all $ x+y+z=1$, $x,y,z\geq 0$. Let us introduce the following function
$$
f(x,y)= 2\bigg(A y(1-x-y)+B x(1-x-y)+C xy\bigg)^{1/2}-\a x-\b y-\g (1-x-y),
$$
where the arguments $(x,y)$ satisfy $0\leq x+y\leq 1$. One can check that this function reaches its maximum on the boundary of $0\leq x+y\leq 1$.
Hence, it is enough to study the following function
$$
g(y)=f(0,y)=2\big(A y(1-y)\big)^{1/2}-\b y-\g (1-y)
$$
on the interval $[0,1]$. One shows that the maximum of $g(y)$ on the interval $[0,1]$ is less or equal than 0 if and only if one has
$ A\leq \b\g$.  Similarly, one finds the other two conditions $ B\leq \a\g$ and $C\leq \a\b$. Hence, if

\begin{eqnarray}\label{q38}
A\leq \b\g, \ \ B\leq \a\g, \ \ C\leq \a\b
\end{eqnarray}
then $\Psi$ is KS-operator. Now, taking into account \eqref{abc} and $|q_k|\leq 1$,  the last conditions \eqref{q38} reduce to
$$
p_1^2+p_2^2+p_3^2\leq 1+2p_1p_2p_3 .
$$

\section{proof of Proposition 1}

Denoting  $Y:=\Phi(\oper)$ one has
$$
(Y^{-1/2}\Phi(X)-Y^{1/2}X)^\dagger ( Y^{-1/2}\Phi(X)-Y^{1/2}X)\geq 0
$$
which implies
$$
\Phi(X)^\dagger Y^{-1}\Phi(X)\geq \Phi(X)^\dagger X +X^\dagger\Phi(X)-X^\dagger YX.
$$
This together with KS-condition yields the assertion.

\section{KS Divisibility}

In this section, we show that the generator $ \mathcal{L}^\sharp$ is dissipative which means that the mapping $\Phi$ in \eqref{LII} satisfies  \eqref{QKS}.

\begin{equation}\label{}
  \mathcal{L}^\sharp_t(X) = \frac 12 \sum_{k=1}^3  \gamma_k(t)( \sigma_k X \sigma_k - X) = \Phi_t(X) - \{ \Phi_t(\oper),X\} ,
\end{equation}
where $\Phi_t(X) =  \frac 12 \sum_{k=1}^3  \gamma_k(t)\sigma_k X \sigma_k$. This genetaror gives rise to KS-divisible evolution iff
\begin{equation}\label{}
   \Phi_t(X^\dagger X) \geq \Phi_t(X^\dagger) X + X^\dagger\Phi_t(X) - X^\dagger \Phi_t(\oper) X .
\end{equation}
To simplify notation we skip time-dependence. Note, that $\Phi(\oper) = \gamma \oper$, with

\begin{equation}\label{gggg}
  \gamma = \frac{1}{2}(\gamma_1+\gamma_2+\gamma_3) .
\end{equation}
Let us observe that the condition \eqref{QKS} can be rewritten as follows:
$$
\Phi(X^\dagger X)-\Phi(X)^\dagger X-X^\dagger \Phi(X)+ \gamma X^\dagger X\geq 0
$$
So, we have
$$
\Phi(X^\dagger X)-  \gamma^{-1} \Phi(X^\dagger) \Phi(X) +  \gamma^{-1} \Phi(X^\dagger) \Phi(X)-\Phi(X^\dagger) X- X^\dagger \Phi(X)+ \gamma X^\dagger X\geq 0 ,
$$
and hence
\begin{equation}\label{}
\Phi(X^\dagger X)- \gamma^{-1}\Phi(X^\dagger)\Phi(X)+[ \gamma^{-1/2}\Phi(X)-  \gamma^{1/2}X]^\dagger [ \gamma^{-1/2}\Phi(X)- \gamma^{1/2}X]\geq 0 .
\end{equation}
Now, to simply analysis without loosing generality  we put $\gamma=1$.  We show that if condition (\ref{g2g}) is satisfied for any $t \geq 0$, then $\Phi$ satisfies

\begin{equation}\label{KS1}
\Phi(X^\dagger X)- \Phi(X^\dagger)\Phi(X)+[\Phi(X) - X]^\dagger [ \Phi(X) - X]\geq 0 .
\end{equation}
One has for $X= w_0 \oper + \wb \cdot \sigma$

\begin{equation}\label{}
\Phi(X)=w_0 \oper + T\wb\cdot\s , \ \ \Phi(X^\dagger) = \overline{w_0} \oper + \big(T\overline{\wb}\big)\cdot\s ,
\end{equation}
where the $3\times 3$ matrix $T_{ij}$ reads $T_{ij} = \lambda_i \delta_{ij}$, and the eigenvalues of the map read

\begin{equation}
 \lambda_1 = \frac{1}{2}(\gamma_1-\gamma_2-\gamma_3) , \ \ \lambda_2 = \frac{1}{2}(-\gamma_1+\gamma_2-\gamma_3) , \ \ \lambda_3 = \frac{1}{2}(-\gamma_1-\gamma_2+\gamma_3) .
\end{equation}
One finds

\begin{eqnarray}\label{ks7}
X^\dagger X &=& (|w_0|^2 + \|\wb\|^2)\oper + (w_0 \overline{\wb} + \overline{w}_0 \wb + i \overline{\wb} \times \wb)\cdot \sigma , \\
\Phi(X^\dagger X) &=& \big(|w_0|^2+\|\wb\|^2\big)\oper +\big(w_0T\overline{\wb}+\overline{w_0}T\wb+iT(\overline{\wb}\times \wb)\big)\cdot\s\\ \label{ks9}
\Phi(X^\dagger)\Phi(X) &=& \big(|w_0|^2+\|T\wb\|^2\big)\oper +\big(w_0T\overline{\wb}+\overline{w_0}T\wb+i T\overline{\wb}\times T\wb \big)\cdot\s
\end{eqnarray}
where $\mathbf{a}\times \mathbf{b}$ stands for the vector product of 3-dimensional vectors. One obtains
\begin{eqnarray*}
\Phi(X^\dagger X)-\Phi(X^\dagger\Phi(X) &=& \big(|w_0|^2-\|T\wb\|^2\big)\oper + i\big(T(\overline{\wb}\times \wb) - T\overline{\wb}\times T\wb \big)\cdot\s ,
\end{eqnarray*}
and
\begin{eqnarray*}
(\Phi(X)-X)^\dagger (\Phi(X)-X) = \big((T\overline{\wb}-\overline{\wb})\cdot\s\big) \big((T{\wb}-{\wb})\cdot\s\big) = \|T{\wb}-{\wb}\|^2\oper  + i\big[T\overline{\wb}-\overline{\wb},T{\wb}-{\wb}\big]\cdot\s .
\end{eqnarray*}
Hence, we find

\begin{eqnarray}\label{KS2}
\Phi(X^\dagger X)-\Phi(X^\dagger)\Phi(X)+[\Phi(X)-X]^\dagger [\Phi(X)-X] = a \oper + {\bf b} \cdot s ,
%\big(|w_0|^2+\|T{\wb}-{\wb}\|^2-\|T\wb\|^2\big)\oper  \nonumber \\
%&&+i\big(T\big[\overline{\wb},\wb\big]-\big[T\overline{\wb},T\wb,\big]+\big[T\overline{\wb}-\overline{\wb},T{\wb}-{\wb}\big]\big)\cdot\s %=\big(\underbrace{|w_0|^2+\|\wb\|^2-(\langle T{\wb},\overline{\wb}\rangle + \langle \wb,T\overline{\wb}\rangle \big)}_A\big)\oper
%+i\bigg(\underbrace{T\big[\overline{\wb},\wb\big]+\big[\overline{\wb},\wb\big] %-\big(\big[T\overline{\wb},\wb,\big]+\big[\overline{\wb},T{\wb}\big]\big)}_{\bf B}\bigg)\cdot\s .
\end{eqnarray}
with

\begin{eqnarray}
% \nonumber % Remove numbering (before each equation)
  a &=& 2 \|\wb\|^2-(\langle T{\wb},\overline{\wb}\rangle + \langle \wb,T\overline{\wb}\rangle \big) \\
  {\bf b} &=& T(\overline{\wb}\times \wb) + \overline{\wb}\times \wb - \big( T\overline{\wb}\times \wb + \overline{\wb}\times T{\wb}\big) .
\end{eqnarray}\label{}
Now, (\ref{KS1}) is equivalent to

\begin{itemize}
  \item $a\geq 0$
  \item $a \geq |{\bf b}|$.
\end{itemize}
%Since neither $a$ nor the vector ${\bf b}$ depend on $w_0$ Without loss of generality, we assume that $\|\wb\|\leq 1$.
Using
$$
\langle T{\wb},\overline{\wb}\rangle= \langle \wb,T\overline{\wb}\rangle=\l_1|w_1|^2+\l_2|w_2|^2+\l_3|w_3|^2
$$
one finds
\begin{eqnarray*}
a = %2\|\wb\|^2-(\langle T{\wb},\overline{\wb}\rangle + \langle \wb,T\overline{\wb}\rangle \big) = 
2\Big[ (1-\l_1)|w_1|^2+(1-\l_2)|w_2|^2+(1-\l_3)|\w_3|^2 \Big] ,
\end{eqnarray*}
and hence taking into account that $\gamma_1+\gamma_2+\gamma_3=2$, one finds

\begin{equation}\label{}
  (\gamma_2+\gamma_3)|w_1|^2+(\gamma_3+\gamma_1)|w_2|^2+(\gamma_1+\gamma_2)|\w_3|^2 \geq 0 ,
\end{equation}
which reproduces condition (\ref{gg}). Hence, condition $a \geq 0$ is equivalent to P-divisibility. The second condition $a \geq |{\bf b}|$ provides further restriction which clearly shows that KS-divisibility implies P-divisibility. One finds for the 3-vector ${\bf b}$:

\begin{equation}\label{}
  {\bf b} = \bigg(\mu_1 (\bar{w}_2 w_3-\bar{w}_3 w_2),\mu_2(\bar{w}_3 w_1 -\bar{w}_1 w_3),\mu_3 (\bar{w}_1 w_2-\bar{w}_2 w_1)\big) = S (\overline{\bf w} \times {\bf w}) ,
\end{equation}
where the $3 \times 3$ matrix $S$ reads $S_{ij} = \mu_i \delta_{ij}$, 
with

\begin{eqnarray} \label{mmm}
&&\mu_1 = 1+\l_1-\l_2-\l_3 = 2 \gamma_1 ,\nonumber \\
&&\mu_2 = 1-\l_1+\l_2-\l_3 = 2 \gamma_2 ,\\
&&\mu_3 = 1-\l_1-\l_2+\l_3 = 2 \gamma_3  \nonumber ,
\end{eqnarray}
where again we took into account $\frac 12(\gamma_1+\gamma_2+\gamma_3)=1$. The condition  $\|{\bf b}\|\leq a$ reads
\begin{eqnarray*}\label{KS3}
 \bigg(\mu_1^2|\bar{w}_2 w_3-\bar{w}_3 w_2|^2+\mu_2^2|\bar{w}_1 w_3-\bar{w}_3 w_1|^2+\mu_3^2|\bar{w}_1 w_2-\bar{w}_2 w_1|^2\bigg)^{1/2} \leq  (1-\l_1)|w_1|^2+(1-\l_2)|w_2|^2+(1-\l_3)|\w_3|^2 .
\end{eqnarray*}
% Without loss of generality, we assume that $\|\wb\|\leq 1$. 
Introducing the following parametrization: $w_1=x e^{i\a_1}$, $w_2=y e^{i\a_2}$, $w_3=z e^{i\a_3}$, with $x,y,z \geq 0$, it reduces to
\begin{eqnarray}\label{KS4}
2\bigg(\mu_1^2 y^2 z^2\sin^2\theta_1 + \mu_2^2 z^2 x^2\sin^2\theta_2 + \mu_3^2 x^2 y^2\sin^2\theta_3\bigg)^{1/2} \leq
(\gamma_2+\gamma_3)x^2+(\gamma_3+\gamma_1)y^2+(\gamma_1+\gamma_2)z^2 ,
\end{eqnarray}
where $\theta_1+\theta_2+\theta_3=2\pi$. The inequality is clearly satisfied if all $\gamma_k\geq 0$. Now, suppose that $\gamma_3 < 0$ (note, that only one $\gamma_k$ can be negative) and let us consider the worst case scenario maximizing LHS and minimizing RHS of (\ref{KS4}). Since

\end{widetext}

$$    \gamma_1+\gamma_2 \geq \gamma_1+\gamma_3 \geq 0 , $$
and

$$   \gamma_1+\gamma_2 \geq \gamma_2+\gamma_3 \geq 0 $$
let us take $z=0$ and $\sin\theta_3=1$. It leads to

\begin{eqnarray}\label{KS8}
|\gamma_3| \sqrt{xy} \leq \frac12\bigg((\gamma_2+\gamma_3)x+(\gamma_1+\gamma_3)y\bigg),
\end{eqnarray}
and hence 

\begin{eqnarray}\label{KS9}
 \sqrt{xy} \leq \frac 12\bigg(\frac{\gamma_2+\gamma_3}{|\gamma_3|} x + \frac{\gamma_1+\gamma_3}{|\gamma_3|} y\bigg),
\end{eqnarray}
for all $x,y \geq 0$. 

\begin{Lemma} Let $\alpha,\beta >0$. Condition

$$   \sqrt{xy} \leq \frac 12\bigg(\alpha x + \beta y\bigg), $$
is satisfied for all $x,y \geq 0$ iff $\alpha\geq 1$ and $\beta\geq 1$. 
\end{Lemma}
Clearly, $\alpha\geq 1$ and $\beta\geq 1$ is sufficient. To show that it is also necessary take $y = 1/x$. It gives

$$   2x \leq \alpha x + \beta/ x . $$
The RHS is minimal for $x = \sqrt{\beta/\alpha}$ and hence

$$  2 \sqrt{\beta/\alpha} \leq 2 \sqrt{\beta\alpha} $$
which give $\alpha \geq 1$. In the same way we prove that $\beta \geq 1$. Summarising, we showed that

$$  \gamma_2+\gamma_3 \geq |\gamma_3| , \ \ \ \gamma_1+\gamma_3 \geq |\gamma_3| , $$
which can be rewritten as 

$$   \gamma_i + 2 \gamma_3 \geq 0 \ , \ \ i=1,2 . $$
Clearly $\gamma_1 + 2\gamma_2 \geq 0$ and $\gamma_2 + 2 \gamma_1 \geq 0$.

%\end{widetext}

\end{document}